\documentclass[11pt]{JHEP3}
\usepackage{graphicx}
\usepackage{epsfig}
\usepackage{amssymb}
\input epsf.tex

\newcommand{\del}{\ensuremath{\partial}}

\newcommand{\half}{\ensuremath{\frac{1}{2}}}

\newcommand{\be}{\begin{equation}}
\newcommand{\ee}{\end{equation}}
\newcommand{\ba}{\begin{eqnarray}}
\newcommand{\ea}{\end{eqnarray}}

\newcommand{\ns}{\normalsize}

\title{Oscillons and quasi-breathers in D+1 dimensions}

\author{
  Paul M. Saffin$^{1}$\footnote{email: paul.saffin@nottingham.ac.uk} and
  Anders Tranberg$^{2}$\footnote{email: a.tranberg@damtp.cam.ac.uk},\\
  $^1${\it\ns School of Physics and Astronomy, University of Nottingham}\\
  {\it\ns University Park, Nottingham NG7 2RD, United Kingdom.}\\  
  $^2${\it\ns DAMTP, University of Cambridge} \\
  {\it \ns  Wilberforce Road, Cambridge, CB3 0WA, United Kingdom.}
}


\preprint{DAMTP-2006-??}

\date{}

\maketitle

\abstract{
We study oscillons in $D+1$ space-time dimensions using a spherically symmetric
ansatz. From Gaussian initial conditions, these evolve by
emitting radiation, approaching ``quasi-breathers'',
near-periodic solutions to the equations of motion. Using a truncated
mode expansion, we numerically determine these quasi-breather
solutions in $2<D<6$ and the energy dependence on the oscillation
frequency. In particular, this energy has a minimum, which in turn
depends on the number of spatial dimensions. We study the time
evolution and lifetimes of the resulting quasi-breathers, and
show how generic oscillons decay into these before disappearing
altogether. We comment on the apparent absence of oscillons for $D>5$ and
the possibility of stable solutions for $D\leq 2$.}

\begin{document}



\section{Introduction\label{sec:introduction}}
Topological defects such as monopoles, vortex strings and domains walls
owe their stability to non-trivial boundary conditions or, equivalently, a conserved
(topological) charge \cite{VS}.
There are other mechanisms which can lead to the formation of lumps in field theory,
such as the conservation of a N\"other charge allowing for Q-balls \cite{Coleman:1985ki} or it may
simply be energetically favourable as in the case of semi-local strings \cite{Vachaspati:1991dz}.

However, it turns out that even in the absence
of such topological constraints or symmetry arguments, solutions to non-linear field
equations can have localised lumps of energy. A typical example of this is the solitonic breather 
solution in the sine-Gordon model. 

A more curious set of lumps in field theories are those which exist 
for a long, but finite, time without any of the
reasons 
listed above.
These go by the name of oscillons 
\cite{Gleiser:1993pt} and exist
as a localized oscillating state, with the oscillations remaining coherent for much longer
than naive dimensional analysis would suggest \cite{Bogolyubsky:1976yu,Bogolyubsky:1976nx}. 
Simulations of $\phi^{4}$ scalar theories
in two dimensions \cite{Hindmarsh:2006ur} suggest that they decay only after millions of natural time units
whereas in three dimensions they last for thousands of natural time units \cite{Copeland:1995fq}. 
The mechanism for their demise follows from their oscillating nature,
with the emitted radiation eventually causing oscillons to decay. 
There is precedence for oscillating
solutions to continue for ever, although it has only been observed in 1+1 dimensions in 
breathers of the sine-Gordon model. However, this model is rather special in that it is integrable, containing
an infinite number of conserved currents. An understanding of the evolution of oscillons was gained in 
\cite{Watkins1996} using Fourier analysis to construct strictly periodic solutions, later termed ``quasi-breathers''
in \cite{Fodor:2006zs}. Such exactly periodic solutions exist because they include a component of inwardly directed
radiation. So while quasi-breathers are not expected to be physical their construction shows that the amount of radiation involved
is small enough (in some cases) to make them a valid approximation. 

In \cite{Gleiser:2004an, Gleiser:2006te}
it has been suggested that oscillons cease to exist, in any meaningful sense, for dimensions greater than 5 or 6
depending on the details of the potential. 
This result was found by assuming a Gaussian profile for oscillons and substituting this into the action in
order to generate an effective action. Given the extreme sensitivity of oscillon lifetimes on the initial
profile \cite{Copeland:1995fq,Honda:2001xg} this approach can only yield an indication of what may happen
and serves as impetus for our study.
We use quasi-breathers in our analysis to
show how oscillons behave in various dimensions, including their non-existence in higher dimensions.

It is worthwhile mentioning at this point that oscillons have been observed in the laboratory  \cite{Umbanhowar:1996},
forming in granular materials placed on a vibrating plate. While these oscillons are not described by a relativistic
field theory they are similar enough to bear the same name.

Apart from the challenge of understanding oscillons in themselves,
interest derives from the possible impact of these objects if they are generated in large numbers 
in phase transitions. In particular, if such ``coherently oscillating'' lumps persist for hundreds 
or thousands of oscillations, they may play a role during post-inflationary preheating and baryogenesis, 
and delay the subsequent equilibration and thermalisation.

Oscillons have been extensively studied in recent years, mostly in
scalar models, but an oscillon has also been reported in the SU(2)-Higgs model \cite{Farhi:2005rz}. 
The very recent work of
\cite{Fodor:2006zs} is a detailed numerical study of oscillons in
3+1 dimensions. Improving on the analysis of \cite{Honda:2001xg},
the authors propose that generic oscillons are composed of a core
quasi-breather of a particular oscillation frequency $\omega$ and a
number of additional modes, depending on the initial state. As time
goes on the oscillon will radiate the excess energy while changing
frequency \cite{Watkins1996}, its core scanning through a series of quasi-breathers,
eventually approaching a critical $\omega$ quasi-breather, after which
the oscillon decays. The critical $\omega$ corresponds to the lowest
energy quasi-breather.

The study in two dimensions of \cite{Hindmarsh:2006ur} showed that a non-spherical perturbation was
not deleterious to the oscillon's evolution and so for our study we restrict our attention to spherical symmetry
in $D$ space dimensions.
The spatial dimension then simply appears as a parameter in an effective 1+1 spacetime field theory,
this allows us to consider non-integer dimensions as a way of observing more clearly how oscillon
behaviour depends on $D$.
Using this ansatz, we will perform a mode-decomposition of the
quasi-breather in a way similar to \cite{Watkins1996,Fodor:2006zs,Honda:2001xg},
and numerically find solutions for each $\omega$, for each choice of
spatial dimensions $D$. 
Restricting to 3+1 dimensions, it was observed in \cite{Watkins1996} that there is a particular value of 
qusi-breather frequency, $\omega_{\rm crit}$, which has minimum energy. Oscillons with frequencies different
from $\omega_{\rm crit}$ evolve toward this solution as they radiate energy, decaying once they reach this
frequency. We shall see that this behaviour persists in other dimensions, although the time it takes
for oscillons to reach $\omega_{\rm crit}$ depends sensitively on $D$. We shall confirm the fact that for $D=3$
lifetimes are of the order of $10^{4}$ \cite{Fodor:2006zs}, whereas for $D=2$ oscillons can
oscillate for many times $10^6$. Obviously, in the limit of $D=0$ an oscillon is simply a particle oscillating
in a well and so is strictly periodic.

These
quasi-breather profiles can then be used as initial conditions for
real-time simulations in order to find the lifetime of these
objects. Also, we will use (non fine-tuned) Gaussian initial
conditions as in \cite{Fodor:2006zs,Honda:2001xg} and study how these
oscillons approach the quasi-breather evolution as radiation is
emitted. We will see that the oscillon closely follows the
quasi-breather, and that the two decay in a similar way. 

In section \ref{sec:model} we will set up the model, the quasi-breather ansatz and the
mode decomposition we use. In section \ref{sec:shooting} we will
determine the quasi-breather solutions numerically, for a range of
$\omega$ and $D$. The real-time evolution of these quasi-breathers is
presented in section \ref{sec:realtime}, where we also compare to the
evolution of a generic Gaussian initial condition. We will discuss the
lifetimes of these objects in section \ref{sec:lifetimes} and conclude
in section \ref{sec:conclusion}.


\section{Model, ansatz and mode expansion\label{sec:model}}

We shall study oscillons in scalar $\phi^{4}$ in $D+1$ dimensions
starting with the action
\ba
S&=&-\int d\hat td^D\hat x\left[ \half\hat \del_\mu\hat\phi\hat\del^\mu\hat\phi+\half m^2\hat\phi^2
                    +\frac{1}{3}\alpha\hat\phi^3+\frac{1}{4}\beta\hat\phi^4\right].
\ea
From this staring point we restrict the number of free parameters by making the potential
a symmetric double well, achieved by imposing
\ba
\alpha&=&-3m\sqrt{\beta/2}.
\ea
This gives a potential with extrema at
\ba
\hat\phi_{ext}=0,\frac{m}{\sqrt{2\beta}},m\sqrt{\frac{2}{\beta}}.
\ea
It is now possible to perform some rescalings of the co-ordinates and
the field
\ba
\hat\phi=\frac{m}{\sqrt{2\beta}}\phi,\quad r=m \hat r,\quad t=m \hat t,
\ea
and consider only radially symmetric field configurations to find
\ba
S&=&-\frac{m^{4-D}}{\beta}A_{D-1}\int dt\;dr\;r^{D-1}\left[ 
                    -\frac{1}{2}(\del_{t}\phi)^2+\frac{1}{2}(\del_r\phi)^2+\half \phi^2
                    -\frac{1}{2}\phi^3+\frac{1}{8}\phi^4\right],
\ea
where $A_{D-1}=2\pi^{D/2}\Gamma(D/2)$ is the volume of the $D-1$ sphere which has been integrated out.

A quasi-breather is a periodic solution to the equations of motion,
and as such can be expanded in modes with frequencies that are
integer multiples of some basic frequency $\omega$, 
\ba
\label{eq:fourierExp}
\phi(r,t)&=&\frac{1}{\sqrt{2}}\phi_0(r)+\sum_{n=1}^{\infty}\phi_n(r)\cos(n\omega t).
\ea
The $\frac{1}{\sqrt{2}}$ in the first term is purely for convenience later.
By integrating the action over a period, $2\pi/\omega$ we have
\ba
\label{eqn:S}
S&=&-\frac{m^{4-D}\pi}{\omega\beta}A_{D-1}\int dr\;r^{D-1}\left[ 
                    \left(\sum_{n=0}^\infty \frac{1}{2}(\del_r\phi_n)^2\right)-U_{eff}(\phi_n)\right],\\
\label{eqn:Ueff}
U_{eff}(\phi_n)&=&-\half\phi_0^2-\half(1-\omega^2)\phi_1^2-\half(1-4\omega^2)\phi_2^2-...\\
               &~&+\half\phi_0^3+\frac{3}{2}\phi_0\phi_1^2
                  -\frac{1}{8}\phi_0^4-\frac{3}{4}\phi_0^2\phi_1^2-\frac{3}{16}\phi_1^4+...~~,
\ea
We now have an effective action in terms of dimensionless parameters $r,\omega$ for the
dimensionless fields $\phi_n$. This allows us to reinterpret it in
terms of a particle rolling in ``time'' $r$ through a
space spanned by the $\underline{\phi}$
coordinates under the influence of the potential (\ref{eqn:Ueff}). 
A contour plot of this potential, setting $\phi_2=\phi_3=...=0$ and taking $\half<\omega<1$
is shown in Fig. \ref{fig:Ueff}, 
with a maximum at $(\phi_0,\phi_1)=(1,\sim 1.3)$, minimum at $(1,0)$ and a saddle at $(0,0)$. We shall use this
plot to undertand the basic shape of the quasi breather profile.

\FIGURE{
\epsfig{file=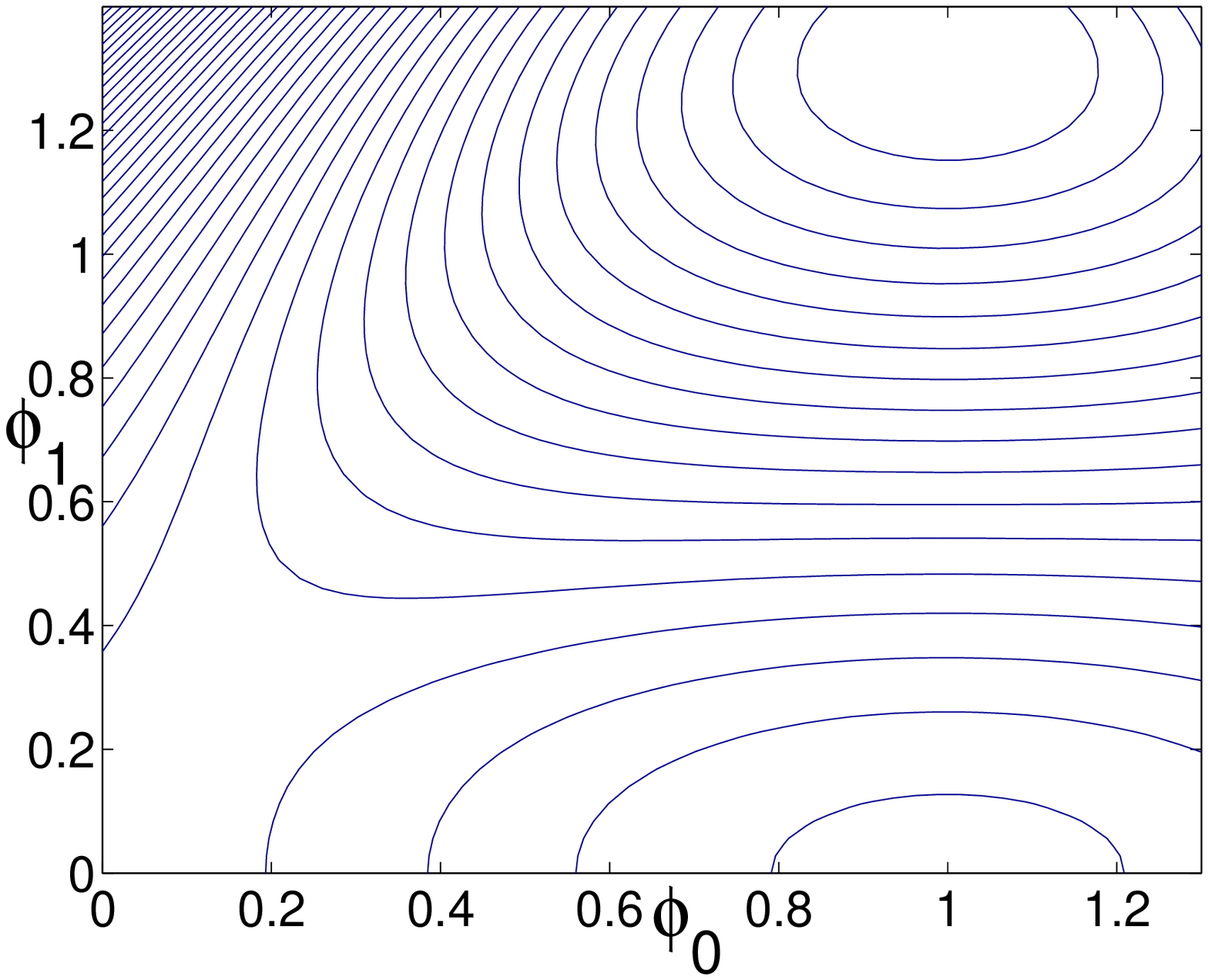,width=7cm,clip}
\caption{A slice through the effective potential. Maxima are at (0,0), (1,1.3) and there
is a minimum at (1,0).}
\label{fig:Ueff}
}


\section{Finding the quasi-breather profile\label{sec:shooting}}

To find the quasi-breather profile for a particular choice of frequency, $\omega$,
we need to solve for the trajectory of a particle moving in $U_{eff}$.
Note however that, because of the $r^{D-1}$ in $S$, there is a frictional force acting to slow down the
particle in analogy to finding the bounce profile in quantum tunnelling \cite{Coleman:1977py}.
Asymptotically, $r\rightarrow\infty$, we require that the particle takes the value $\underline\phi=\underline 0$,
corresponding to the scalar field living in the vacuum asymptotically. 
If we were to ignore those $\phi_n$ with $n\geq 2$ the problem would be to find the location in Fig. \ref{fig:Ueff}
where a particle can start at ``time'' $r=0$ in order to end at $\underline\phi=\underline 0$ as $r\rightarrow\infty$.
This is achieved by taking an initial guess at the starting point and observing whether the particle undershoots
or overshoots $\underline\phi=\underline 0$, then amending the starting point appropriately. In the actual problem
one cannot neglect $\phi_{n\geq 2}$ but the process for finding the values $\phi_0(0)$ and $\phi_1(0)$ is the same.
The starting values of $\phi_{n\geq 2}$ is less critical, as can be understood from (\ref{eqn:Ueff}), and correspond
to adjusting the amount of radiation contained in the quasi-breather solution.
Near $\underline\phi=\underline 0$, i.e. the asymptotic region,
we see that the effective potential is quadratic
with $\phi_n$ aquiring a mass-squared of $-\half(1-n^2\omega^2)$, so those $\phi_n$ with $n>1/\omega$
have positive mass-square while those with $n<1/\omega$ have negative mass-square. This means that the
$\phi_{n>1/\omega}$ are oscillatory in the asymptotic region, corresponding to radiation. 
We shall only concern ourselves with $\half<\omega<1$ so that $\phi_{n\geq 2}$ correspond to radiation modes.
With (\ref{eq:fourierExp}) being an expansion of spatially oscillating
standing waves, this displays the fact that there is
ingoing radiation balancing the outgoing.

\subsection{Shooting\label{sec:shootingalgorithm}}

\FIGURE{
\epsfig{file=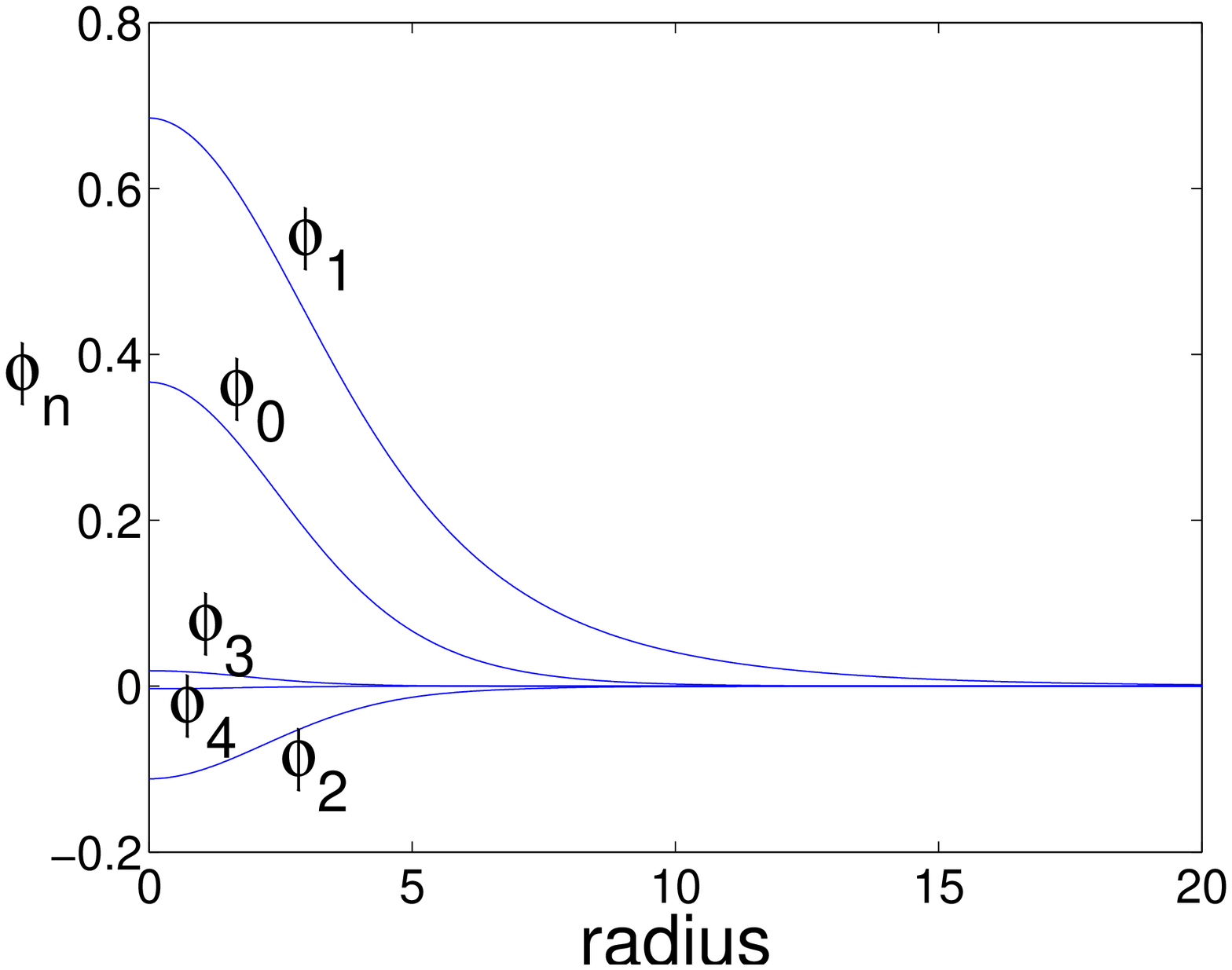,width=7cm,clip}
\epsfig{file=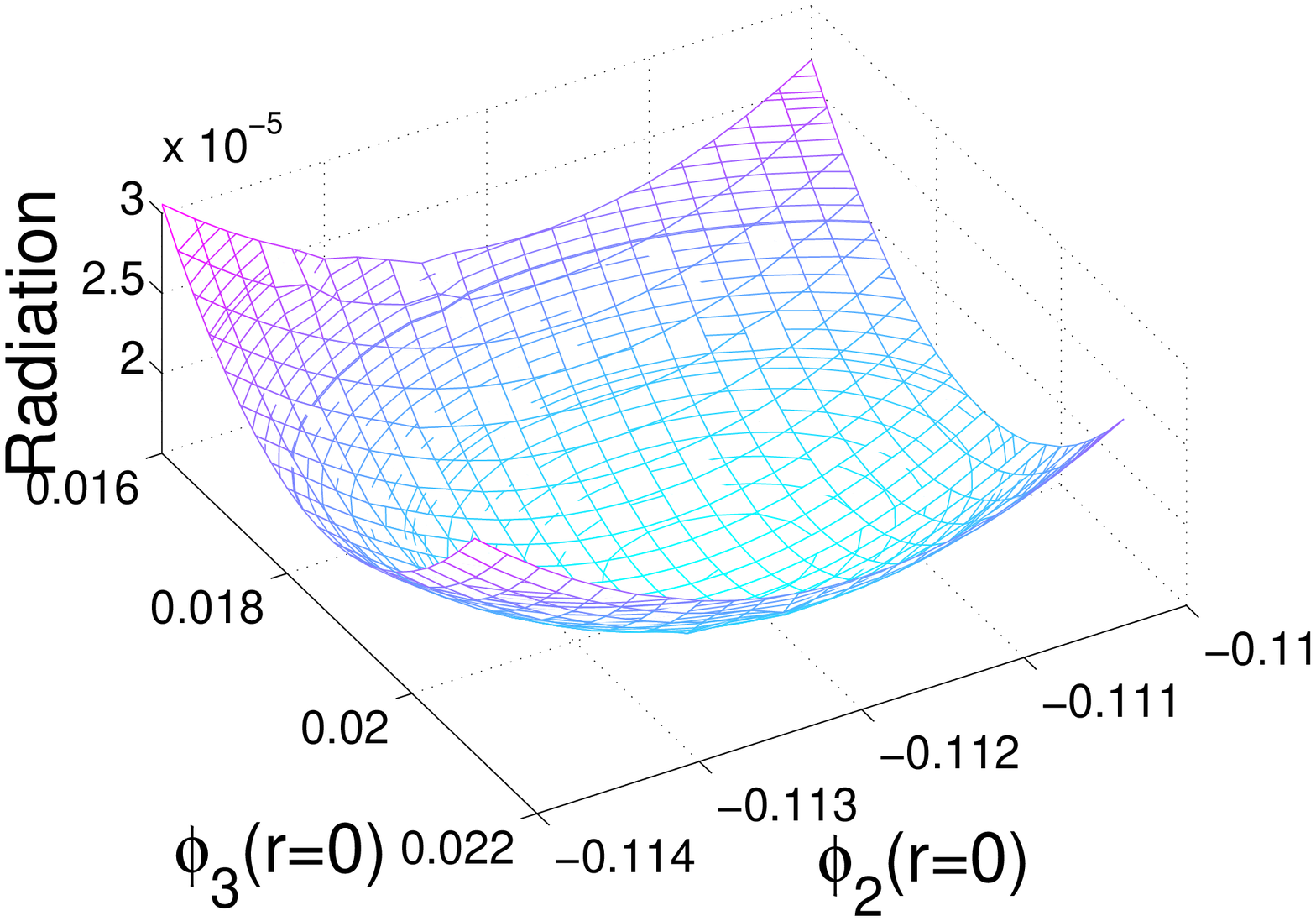,width=7cm,clip}
\caption{Left: The profiles of $\phi_0$, $\phi_1$, $\phi_2$, $\phi_3$
  and $\phi_4$ for $D=3$ and $\omega=0.97$. Right: The dependence of the quasi-breather radiation 
  on $\phi_2(r=0)$ and $\phi_3(r=0)$.}
\label{fig:profiles5}
}

A general analytical solution of the equation of motion does not
exist, and we will instead proceed to find the quasi-breather profile by
numerical means, using a shooting algorithm. Although one can in principle shoot
with multiple degrees of freedom we will truncate the mode expansion, using
the first four terms, $\phi_{n}$, $n=$0, 1, 2, 3 in our calculations of energy
and radiation. Once we have a profile with these components we include the
fifth field, $\phi_{4}$, as a consistency check to make sure that it does indeed
form only a small contribution.

In the shooting algorithm we pick initial values for $\phi_{n}$, $n<4$ at $r=0$, requiring
$\phi_n'(r=0)=0$ for regularity. As mentioned earlier, the values of the radiation modes,
$n=$2, 3, simply dictate how much radiation the quasi breather contains. In \cite{Fodor:2006zs}
the solution is taken 
which minimizes the amplitude of $\phi_2$
asymptotically, here we aim to find the solution which minimizes the total radiation as discussed
below. For any particular choice of $\phi_2(0)$ and $\phi_3(0)$ we then vary $\phi_0(0)$ and $\phi_1(0)$
until we find the solution which approaches $\underline\phi=0$ asymptotically. This then gives us the quasi
breather for frequency $\omega$ and the chosen $\phi_2(0)$, $\phi_3(0)$. By finding the quasi breathers
for a range of $\phi_2(0)$, $\phi_3(0)$ we are able to select the one which has the minimum amount of radiation.

An example of a set of profile functions is shown in
Fig. \ref{fig:profiles5} (left); one can see that the amplitudes of the high-$n$ modes decreases, allowing
for a valid truncation approximation.
As we vary the values of $\phi_2(r=0)$ and $\phi_3(r=0)$ in the shooting algorithm we find quasi-breathers
with differing amounts of radiation, with one choice giving a minimum, Fig. \ref{fig:profiles5} (right). 
We take this solution with minimum radiation to be our quasi-breather.

\subsection{Radiation energy\label{sec:radiation}}

As the oscillon is made by balancing incoming and outgoing radiation then in
the realistic situation where there is only outgoing radiation the oscillons
will necessarily decay. Here we calculate the rate of energy loss as estimated
by constructing a quasi-breather.

First we introduce the standard energy momentum tensor for a real scalar field,
\ba
T_{\mu\nu}&=&\del_\mu\phi\del_\nu\phi-g_{\mu\nu}\left[\half\del^\rho\phi\del_\rho\phi+V\right],\nonumber\\
\ea
then we define the momentum flux
\ba
P^\mu&=&T^{0\mu},
\ea
and conservation of the energy-momentum tensor tells us that $P^\mu$ is divergence
free. By taking $P^0$ as the energy density we find
\ba
\dot{E}&=&-\int d^{D}x\sqrt{g_{(D)}}\nabla_i P^i,
\ea
and for a spherically symmetric field we find
\ba
\dot E(R)&=&-A_{(D-1)}R^{D-1}P^r(r=R),
\ea
where we have defined $E(R)$ to be the energy within a radius $R$.

Asymptotically the fields are small so we may approximate the field equation by the Klein-Gordon
equation
\ba
-\ddot\phi+\phi''+\frac{D-1}{r}\phi'&=&\phi,
\ea
the closed form solution of which can be approximated asymptotically
by the form
\ba
\phi\sim \cos(kr-\Omega t+\varphi)/r^{(D-1)/2},\qquad
k^2=\Omega^2-1.
\ea
where we have allowed for a possible phase, $\varphi$.
Our solution is a sum of waves of different frequencies producing a standing wave between ingoing
and outgoing waves,
\ba
\phi(r\rightarrow\infty)&\rightarrow&\frac{1}{r^{(D-1)/2}}\left( f_2\cos(k_2r+\varphi_2)\cos(2\omega t)
                                                               +f_3\cos(k_3r+\varphi_3)\cos(3\omega t)+...\right),\\
&=&\frac{1}{2r^{(D-1)/2}}(\;\;f_2\cos(k_2r+\varphi_2+2\omega t)+f_2\cos(k_2r+\varphi_2-2\omega t)\\\nonumber
&~&\hspace{1.8cm}           +f_3\cos(k_3r+\varphi_3+3\omega t)+f_3\cos(k_3r+\varphi_3-3\omega t) +...),\\
&=&\phi_{in}+\phi_{out},
\ea
where $k^2_n=n^2\omega^2-1$. Using
\ba
P^{r}&=&\dot\phi_{out}\phi_{out}'
\ea
we can integrate over one period and find that the average rate of energy loss over a period is
\ba
\label{eq:radiation}
<\dot E>&=&-\frac{\omega A_{(D-1)}}{8}(2k_2f_2^2+3k_3f_3^2+...).
\ea
For the profiles used in Fig. \ref{fig:profiles5} we find
\ba
\dot E&=&7.9\times 10^{-6}+0.31\times 10^{-6}+0.035\times 10^{-6}+...
\ea
The leading term is from the $\phi_2$, with the $\phi_3$ and $\phi_4$ contributions
being sub-dominant.

\subsection{Quasi-breather results \label{sec:results}}

\FIGURE{
\epsfig{file=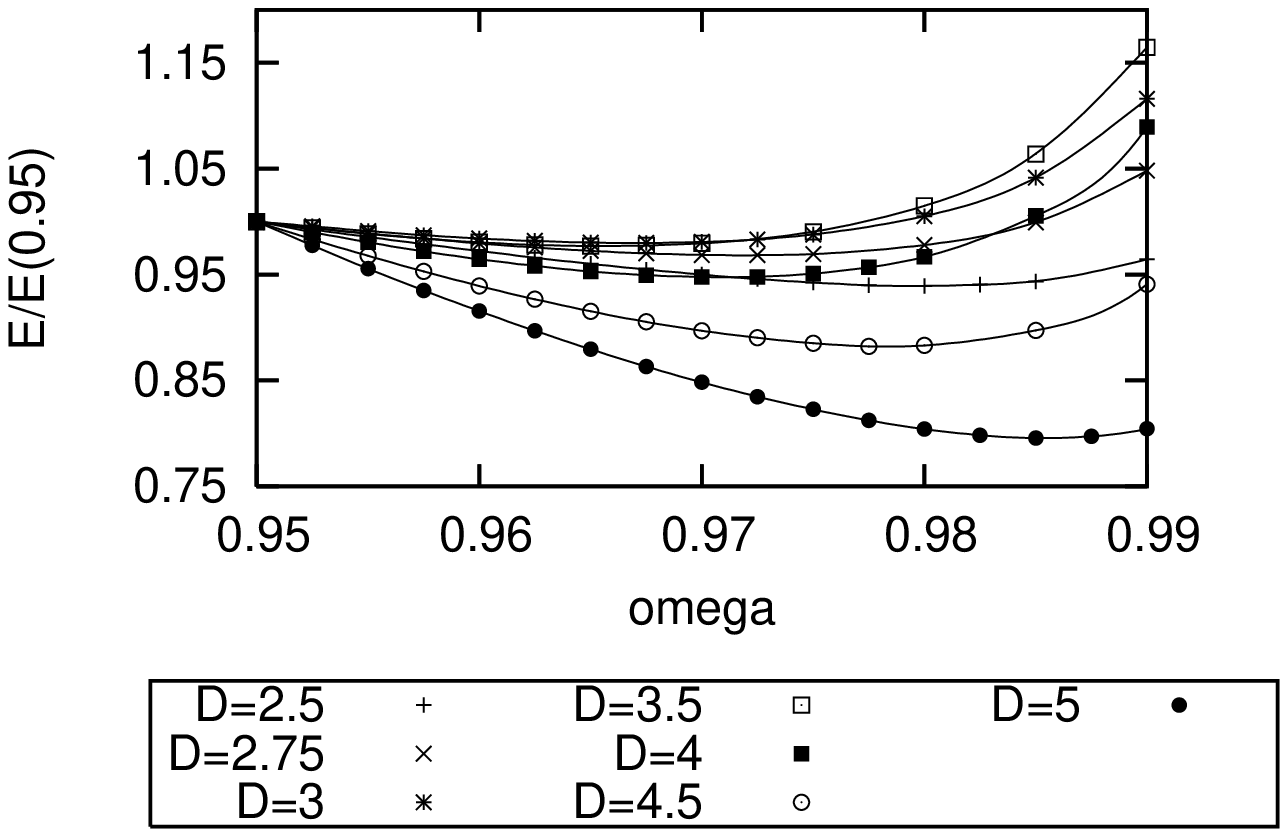,width=8cm,angle=-90,clip}
\caption{The rescaled energy plot.}
\label{fig:Erescaled}
}
\FIGURE{
\epsfig{file=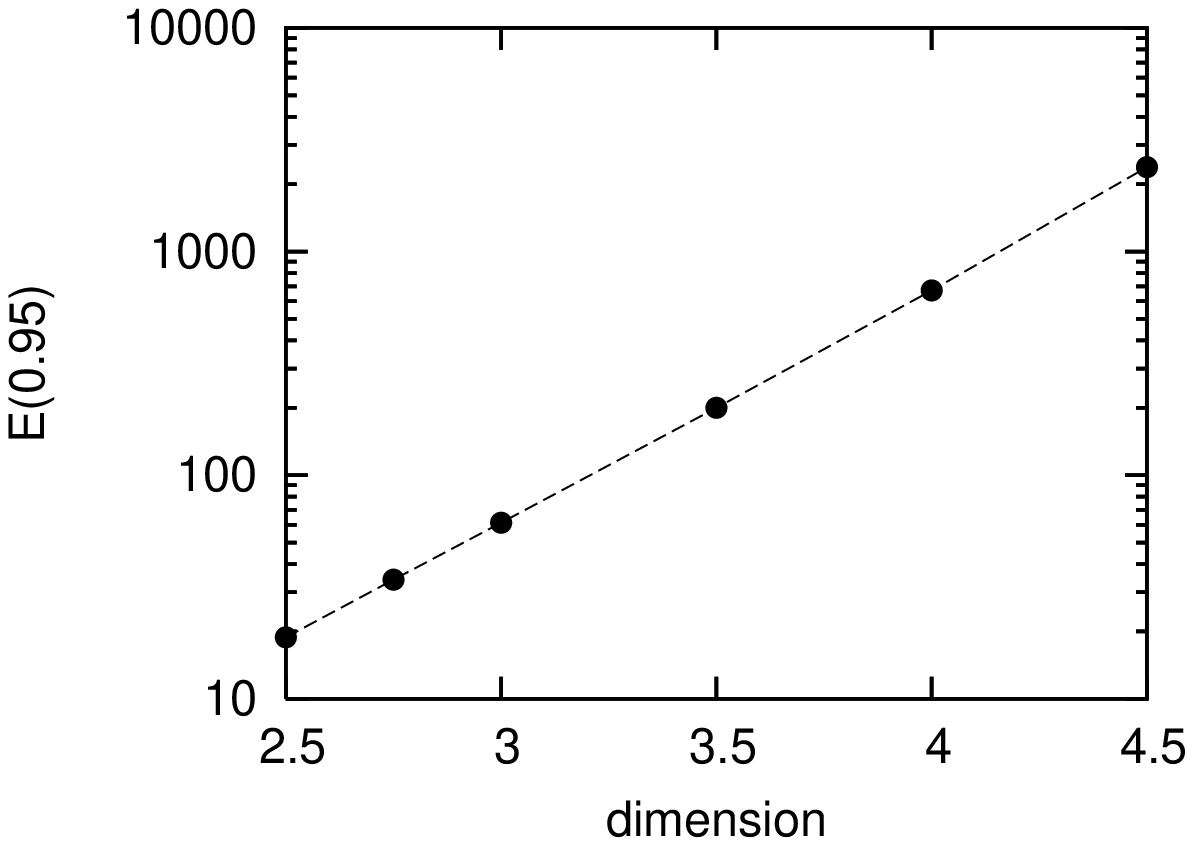,width=8cm,angle=-90,clip}
\caption{$E(D,\omega=0.95)$.}
\label{fig:rescaleFactor}
}

\FIGURE{
\epsfig{file=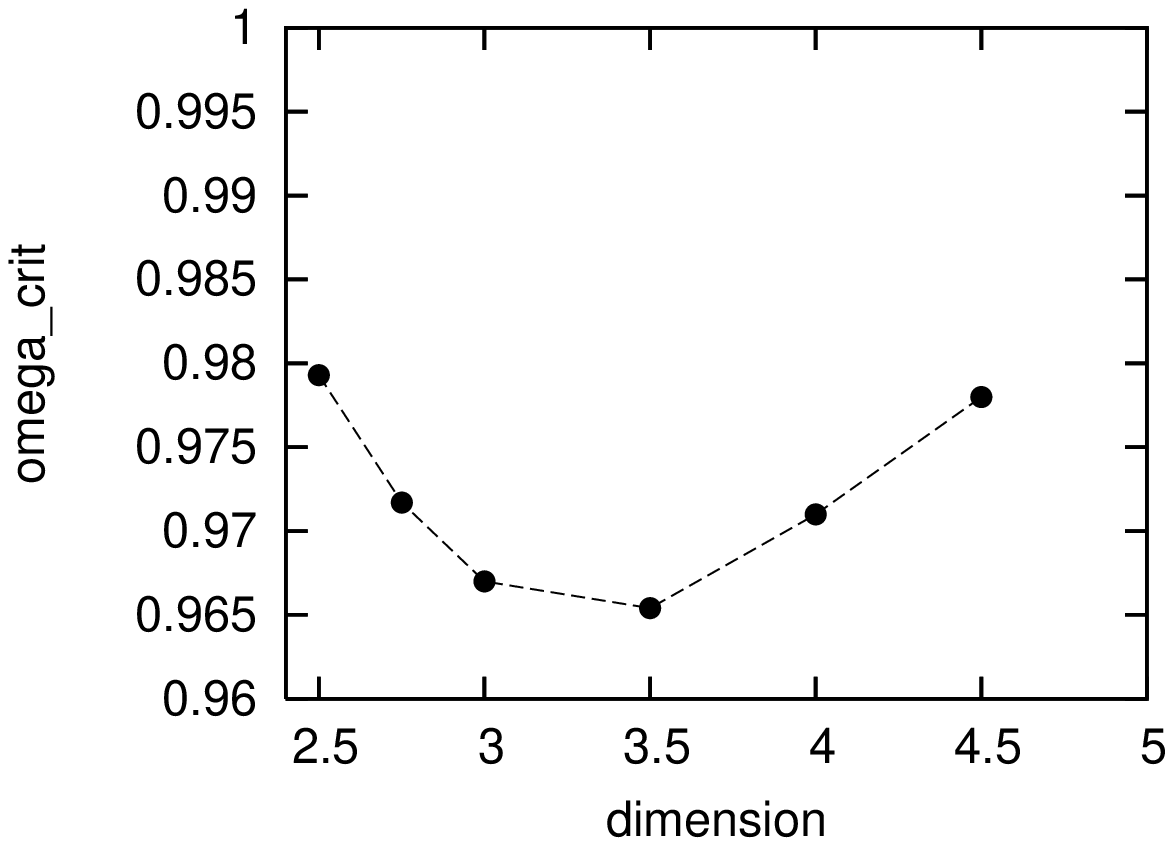,width=8cm,angle=-90,clip}
\caption{The location of the energy minimum.}
\label{fig:energyMinima}
}

\FIGURE{
\epsfig{file=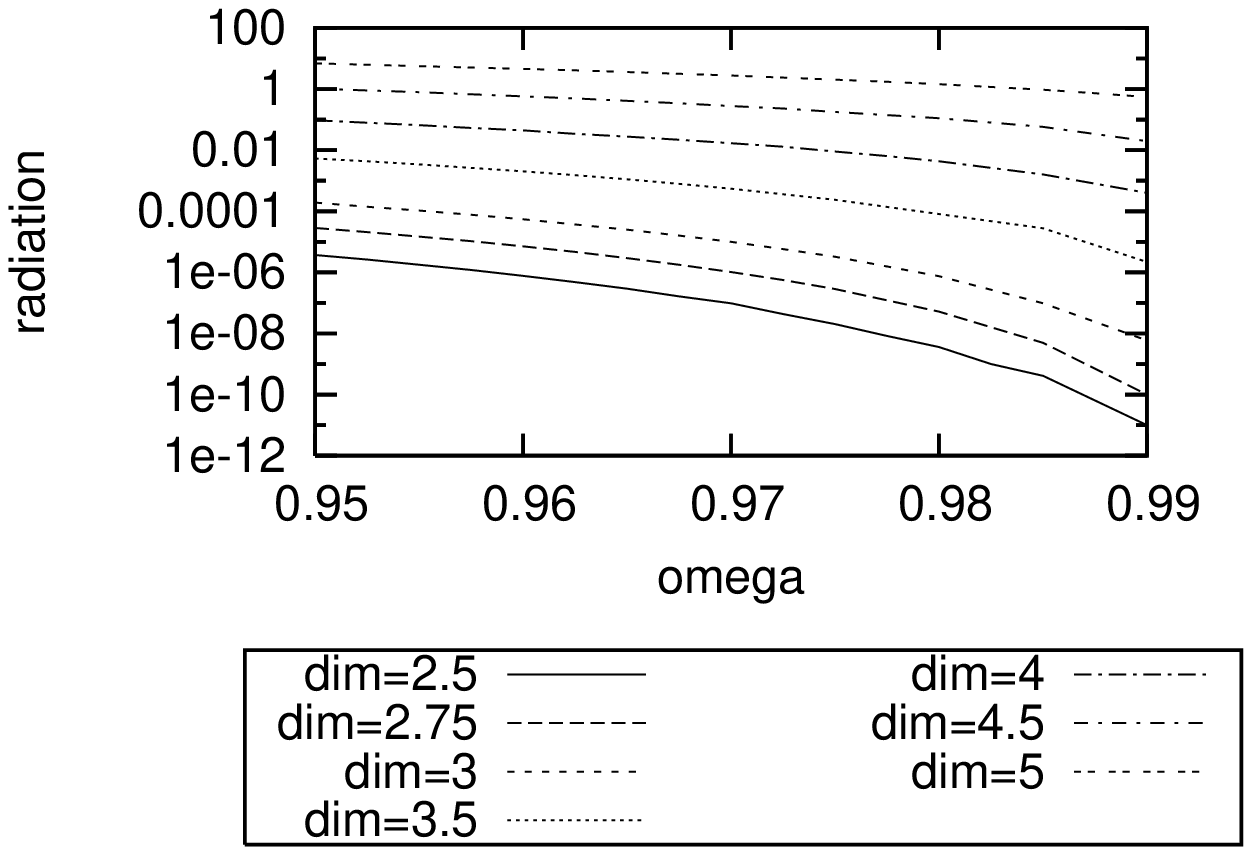,width=8cm,angle=-90,clip}
\caption{The radiation of a quasi breather.}
\label{fig:radiation}
}

\FIGURE{
\epsfig{file=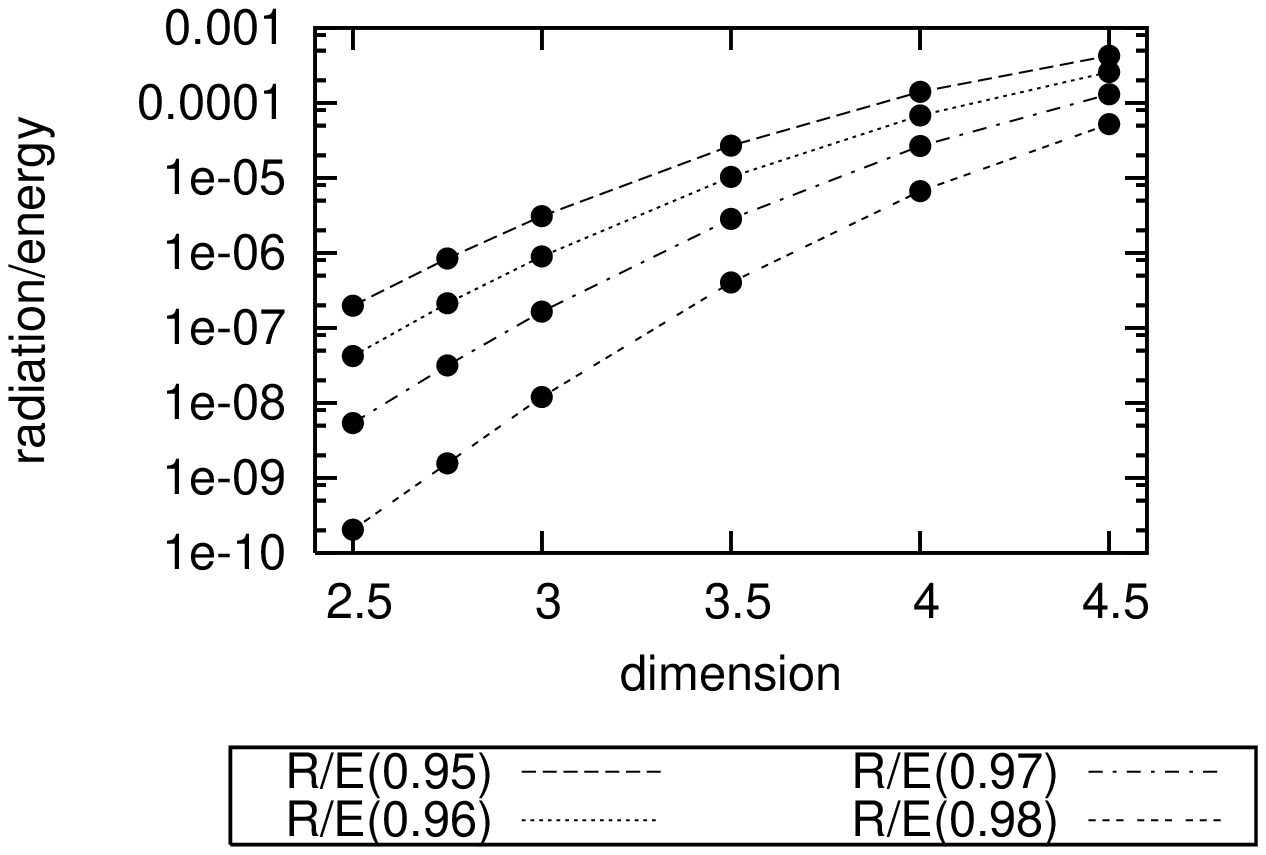,width=8cm,angle=-90,clip}
\caption{The ratio of radiation to energy as a function of dimension.}
\label{fig:RoverE}
}

We now have everything in place to find a quasi-breather for each given frequency, we shall only focus on
frequencies in the range $0.95\leq\omega\leq 0.99$. The choice of lower bound is because, as we shall
see, oscillons spend most of their time above this frequency. Our upper bound is due to numerical
inaccuracies in the shooting algorithm giving unreliable results above this frequency.
The first result to discuss is the energy of the quasi-breathers as $\omega$ varies, then we can understand how
oscillons evolve.

In Fig. \ref{fig:Erescaled} we present a plot, for various dimensions, of the quasi-breather energy as a function
of $\omega$. In order to get the data for the different dimensions on the same axes we have rescaled 
each curve so that they meet at $\omega=0.95$. The different rescaling parameters, $E(D,\omega=0.95)$,
are given in Fig. \ref{fig:rescaleFactor} showing that to a good approximation the energy of a quasi breather
depends exponentially on dimension. In order to calculate the energy of the quasi-breather we note
that the asymptotic form of the field causes the energy, $E(R)$, to diverge linearly with $R$. This divergence
is simply due to the radiation, so to define the energy of the quasi-breather we simply subtract this contribution.

From Fig. \ref{fig:Erescaled} we see that for each dimension considered the energy shows a minimum
at $\omega_{\rm crit}$ as $\omega$ is varied, with $\omega_{\rm crit}$ being dimension dependent. This minimum was first
observed in \cite{Watkins1996} and later in \cite{Fodor:2006zs}. 
We can see how this minimum varies with dimension in Fig. \ref{fig:energyMinima}. Unfortunately our numerical methods
break down below $D\sim 2.5$, because the $\phi_4$ mode becomes relevant,
and above $D\sim 4.5$, because we could not shoot $\underline\phi$ to the vacuum,
so we are unable to give a complete plot of the energy
minima. It is possible that the techniques advocated in \cite{Fodor:2006zs} would give a more complete picture,
it would be particularly interesting to see how the curve develops below $D=2.5$.

As explained in  \cite{Watkins1996} oscillons are expected to evolve by radiating energy, and so move on to a
quasi breather with lower energy and different $\omega$. Once the quasi-breather with $\omega_{\rm crit}$ is reached there is
no quasi-breather with lower energy, so the oscillon radiation causes the lump to radiate away completely. We shall see in the
next section that this is a good description of the dynamics of oscillons, although we remind the reader that
quasi-breathers are only expected to be an approximation to oscillon evolution owing to the presence of ingoing radiation
in a quasi-breather but not in an oscillon.

Along with the calculation of energy for the quasi-breathers we are able to calculate the radiation emitted (and received)
as a function of $\omega$ for each dimension, this is presented in Fig. \ref{fig:radiation}.
A measure of how long an oscillon can be expected to last comes
from the ratio of the rate of energy loss (\ref{eq:radiation}) to energy and is plotted in Fig. \ref{fig:RoverE}.
From this graph we see that as we increase the number of spatial dimensions the amount of radiation relative
to energy inside a quasi-breather increases by many orders of magnitude,
so reducing the lifetime of an oscillon. In fact, we can derive a more complete version of the history of oscillons
using quasi-breathers as we shall see in section \ref{sec:QB_O}.


\section{Real-time evolution\label{sec:realtime}}

Through a detailed minimisation process, we have now determined the solutions
with the lowest radiation for a range of $\omega$ and $D$. The
radiation modes correspond to in- and out-going radiation, and if they
are both included, the quasi-breather should in principle be stable,
possibly up to the order of truncation of the mode expansion.

In physical situations we cannot expect there to be incoming radiation so quasi-breathers
are necessarily an approximation to oscillons. Here we test the state of this approximation
by comparing the real-time evolution of lumps with initial conditions that are either Gaussian
or quasi-breather. The quasi-breather profile would, of course, lead to an exactly periodic evolution so to make the
comparison more sensible we damp the radiation at some large radius.

Ultimately, we are interested in lumps of energy created through
some finite energy process, such as a phase transition, and so a
real-life oscillon is not a quasi-breather. Following \cite{Fodor:2006zs}, we will
distinguish between quasi-breathers and oscillons, and
represent the latter by Gaussian initial conditions.

We will study the time evolution by discretising the equation of
motion, in a simple way. The equation of motion reads, imposing spherical
symmetry explicitly,
\ba
\partial_{t}\phi_{r}(t)+\epsilon\Delta(\phi_{r}(t))
&=&\Pi_{r}(t),\nonumber\\
\partial_{t}\Pi_{r}(t)+\epsilon\Delta(\Pi_{r}(t))
&=&-\Gamma\Pi_{r}(t)
   +\partial_{r}\partial_{r}^{'}\phi_{r}(t)
-\frac{1}{dr}\left(1-(1-dr/r)^{D-1}\right)
  \partial_{r}^{'}\phi_{r}(t)\nonumber\\
&&-\left(1-\frac{3}{2}\phi_{r}(t)+\frac{1}{2}\phi^{2}_{r}(t)\right)\phi_{r}(t),
\ea
with 
\ba
\partial_{t}(...)_{r}=\frac{(...)(r,t+dt)-(...)(r,t)}{dt},\quad
\partial^{'}_{t}(...)_{r}=\frac{(...)(r,t)-(...)(r,t-dt)}{dt},
\ea
and similarly for spatial derivatives, and
\ba
\epsilon\Delta(...)&=&\epsilon \frac{dr^{3}}{dt}\left(\partial_{r}\partial_{r}^{'}\right)^{2}(...).
\ea
We apply this simple leap-frog algorithm with a large lattice (20001 sites), small
lattice spacing ($dr=0.01$), time-step ($dt=0.005$) and an
equally spaced grid with boundary conditions $\phi(r=20001)=0$,
$\partial_{r}\phi(r=0)=0$. In \cite{Fodor:2006zs} it was found that at the
level of fine-tuning applied there, the maximum lifetime and the
corresponding initial condition depend somewhat on the
discretisation. However at the level of precision employed here, this
will not be important; we are not attempting to fine-tune to the maximum lifetime. 

In order to get rid of the emitted radiation, we mimic an infinitely
distant spatial boundary by adding a damping term only
beyond a certain radius $R_{b}$.
\be
\Gamma=0.1,~~~r>R_{b};\quad \Gamma=0,~~~r<R_{b}.
\ee
The oscillon core roughly stretches
to $r=30$ and we concluded that $R_{b}=60$ and $\Gamma=0.1$ were
reasonable choices. In order to get rid of spurious radiation
production on the lattice scale, we also included a Kreiss-Oliger
$\partial_{r}^{4}$ damping term in the equations of motion, controlled by a multiplicative
parameter $\epsilon$ \cite{Honda:2001xg,KO}.

The initial condition for the oscillon is a Gaussian,
\be
\phi(0)_{r} = A \exp{\left(-\frac{r^{2}}{2\sigma^{2}}\right)},
\ee
which is a two-parameter family of profiles, determined by the amplitude $A$ and
the width $\sigma$. We will fix $A=2$, i.e. the center of the oscillon
starts out in the potential minimum at $\phi=2$, while the asymptotic region is in the minimum
at $\phi=0$. 
We will also fix
$\sigma=3.3$, and consider it a generic initial condition. Not all
values of $\sigma$ give long-lived oscillons, but 3.3 is by no means fine-tuned.

We will compare the evolution of the Gaussian oscillon to the trajectory
starting from a quasi-breather.


\subsection{Quasi-breathers and oscillons\label{sec:quasioscillons}}
\label{sec:QB_O}

\FIGURE{
\epsfig{file=./pictures/D3omega_new.eps,width=9cm,clip}
\caption{The effective oscillation frequency for a $D=3$
oscillon with Gaussian initial conditions (black, large oscillations), quasi-breather $\omega=0.95$ initial conditions 
(red/grey) and the
estimate (\ref{eq:t_estimate}) (blue dashed). The orange dotted line represents $\omega_{\rm crit}$.}
\label{fig:omega3D}
}

\FIGURE{
\epsfig{file=./pictures/D3envelope_new.eps,width=7cm,clip}
\epsfig{file=./pictures/D3energy_new.eps,width=7cm,clip}
\caption{The envelope of the center ($r=0$) oscillation (left) and the
  energy within a shell of $r=R_{b}=60$ (right). Black lines is the
  oscillon with Gaussian initial conditions, red/grey using the $\omega=0.95$ quasi-breather
  as an initial condition.}
\label{fig:energyD3}
}

During the real-time evolution the oscillon is not stricly periodic, nevertheless
we can define an effective frequency of the oscillation by studying the core of the oscillon
and defining
\ba
\omega_{\rm eff}(t)=\frac{2\pi}{\delta t_{\rm crossing}},
\ea
where $\delta t_{\rm crossing}$ is simply the time between one crossing of
$\phi(r=0)$ through zero and the second crossing after that. 
This will enable us to measure how the frequency of an oscillon changes in time, as well
as finding the frequency at which the oscillon finally decays.
Both of these quantities can also be calculated within the quasi-breather picture. The
time dependence of $\omega$ is found because
given the energy $E(\omega)$ (Fig. \ref{fig:Erescaled})
we can calculate $dE/d\omega$, so along with (\ref{eq:radiation})
we find $t(\omega)$ to be
\ba
t-t_{i}=\int_{t_{i}}^{t}dt=\int_{\omega_{i}}^{\omega}\frac{dE/d\Omega}{dE/dt}d\Omega.
\label{eq:t_estimate}
\ea
Fig. \ref{fig:omega3D} presents a comparison in three spatial dimensions of $\omega(t)$
for our three cases: the semi-analytic prediction of \ref{eq:t_estimate} (blue, dashed); the
numerical evolution starting with (non fine tuned) Gaussian initial conditions (black);
and the numerical evolution starting with the $\omega_i=0.95$ quasi-breather profile
(red/grey).

The frequency $\omega_{eff}$ for the quasi-breather initial conditions slowly scans through $\omega$ until it
reaches $\omega_{\rm crit}\simeq 0.967$ where it decays; this
happens around $t=6150$. Eq. (\ref{eq:t_estimate}) predicts a longer lifetime by a factor of $\sim3$
giving $t \sim 17000$, this over-estimation turns out to be typical and presumably indicates that the quasi-breather
picture underestimates the amount of radiation. After the decay,
the characteristic frequency becomes $1$, the radiation frequency.

The oscillon from Gaussian initial conditions
starts out far away from the quasi-breather (the center
is initially at $\phi=2$), but proceeds to shed
much of its energy and approach the quasi-breather behaviour. The
frequency has a large modulation, but as the critical frequency is
reached the oscillon decays in a way remarkably similar to the quasi-breather.

This approach of the case with Gaussian initial conditions to that of the quasi-breather
evolution can also be seen by measuring the time development of the core in the
lumps, Fig. \ref{fig:energyD3} (left) shows the
envelope of the time evolution for the core of the quasi-breather (red/grey) and
the Gaussian oscillon (black). The modulation of the oscillon frequency is
manifest, but once the decay happens, the two evolve very similarly.

Finally, we plot in Fig. \ref{fig:energyD3} (right) the energy within a sphere of radius $R_{b}$, for the
Gaussian oscillon (black) and the quasi-breather (red/grey). As expected, the
oscillon sheds its energy out of the ball into radiation (which is
then damped away outside the ball). The energy approaches the
quasi-breather value, and reaches it just in time for the decay. Note
that the decay is delayed by about $600$ in time, compared to the
crossing of the critical frequency.

Recalling that the initial conditions for the Gaussian were not fine-tuned, this is
strong evidence that the quasi-breather behaves as an attractor and is a useful way
of understanding oscillon dynamics.

\subsection{More than 3 dimensions}

\FIGURE{
\epsfig{file=./pictures/D4omega_new.eps,width=7cm,clip}
\epsfig{file=./pictures/D4envelope_new.eps,width=7cm,clip}
\caption{Effective frequency (left) and center envelope (right), for
  $D=4$. Color
  coding as in Fig. \ref{fig:omega3D}.}
\label{fig:omegaD4}
}

As argued in \cite{Gleiser:2004an,Gleiser:2006te} and indicated in Fig. \ref{fig:RoverE} we should expect
that oscillons have rather shorter lives in higher dimensions. The impact of this on the quasi-breather
picture is that we should expect it to be less accurate, after all, the quasi-breathers are strictly
periodic and have infinite lifetimes. Repeating the analysis of the previous section, but now in four
spatial dimensions, we find the results presented in Fig. \ref{fig:omegaD4}. Again we note that the lifetime as predicted
by the quasi-breather arguments is longer than the measured lifetime, by a factor of $\sim 2$ (Fig. \ref{fig:omegaD4} (left)).
However, by observing the evolution
of the core (Fig. \ref{fig:omegaD4} (right)) we see that the oscillon phase ends at around $t\sim 400$
which is where $\omega_{\rm eff}$ hits $\omega_{\rm crit}\sim 0.971$, thus strengthening the argument that oscillons
decay once they reach the critical frequency.

\subsection{Less than 3 dimensions}

\FIGURE{
\epsfig{file=./pictures/D2.75omega_new.eps,width=7cm,clip}
\epsfig{file=./pictures/D2.5omega_new.eps,width=7cm,clip}
\caption{Effective frequency for $D=2.75$ (left) and $D=2.5$ (right). Quasi-breather initial conditions (red/grey),
  Gaussian initial conditions (black), estimate (blue dashed) and critical frequency (orange dotted).}
\label{fig:omegaD2.75}
}

Because of concerns about the validity of the mode truncation, we did
not trust finding quasi-breathers in $D=1,2$; the techniques of \cite{Fodor:2006zs}
may well be more reliable in this regime.
Instead we generalised to non-integer $D$, $D=2.5,2.75$ which allows us to see more clearly
any dependence on spatial dimension. We are able to do this becuase we have assumed spherical
symmetry, meaning that the dimension $D$ appears only as a parameter in our evolution equations.

In Fig. \ref{fig:omegaD2.75} we see the evolution of
$\omega$ for oscillons with quasi-breather initial conditions, Gaussian
initial conditions and the semi-analytic estimate. As expected we see that oscillons last longer
in these smaller spatial dimensions, with the semi-analytic reasoning overestimating the lifetime
by a factor of a few.

Fig. \ref{fig:omegaD2.75} (right) shows
$\omega_{\rm eff}(t)$ for $D=2.5$. Note that the timescale of interest
is now $10^{6}$ and we did not evolve the simulation up to the decay time in this
case; $\omega_{\rm crit}\simeq 0.98$ is far from reached. The Gaussian oscillon
(black) closely follows the quasi-breather oscillon from times $\simeq 20000$.
The estimate of the evolution (blue dashed) is somewhat off in the same way as the other cases, although the
timescale and general trend is well reproduced. The prediction is that
the oscillons could live as long as $8\times
10^{6}$ in natural time units.

\subsection{Approaching $\omega_{\rm crit}$ from above}

\FIGURE{
\epsfig{file=./pictures/omega_largerallD.eps,width=12cm,clip}
\caption{The effective frequency for various $D$ when starting from an
  $\omega=0.99$ quasi-breather. The orange dotted line is $\omega_{\rm
  crit}$.}
\label{fig:omega_above}
}

So far we have considered only those oscillons whose frequency is less than the critical
frequency, and we have seen that as they evolve they follow along one of the curves in
Fig. \ref{fig:Erescaled} from some $\omega<\omega_{\rm crit}$ until they reach $\omega=\omega_{\rm crit}$
at which point they decay. The evolution of the oscillon as it approaches $\omega_{\rm crit}$
is observed to be rather gentle, slowing down because the radiation of a quasi-breather decreases as we approach
$\omega_{\rm crit}$ from below. Fig. \ref{fig:Erescaled} reveals another possibility for oscillons, namely
those starting from $\omega>\omega_{\rm crit}$, in this case the quasi-breather picture would say that $\omega$ 
approaches
$\omega_{\rm crit}$ from above with the oscillon following down one of the curves in Fig. \ref{fig:Erescaled}.
Using Gaussian initial conditions we were not able to find an oscillon with frequency larger than the critical value,
however we are able to start with a quasi-breather profile satisfying $\omega>\omega_{\rm crit}$. The results are 
somewhat
disappointing for the quasi-breather picture, as the subsequent evolution did not show any oscillon phase. Instead, 
with the exception of $D=3$, the field dissipated rapidly straight into radiation, $\omega=1$, as shown in
Fig. \ref{fig:omega_above}.


\section{Oscillon life-times and dimensionality\label{sec:lifetimes}}

\FIGURE{
\epsfig{file=./pictures/intDall.eps,width=7cm,clip}
\epsfig{file=./pictures/lifetimes.eps,width=7cm,clip}
\caption{Left: The evolution of $\omega$ as predicted by the quasi-breather model using
  \ref{eq:t_estimate}. Right: Lifetimes for an oscillon starting from a Gaussian with
  $\sigma=3.3$ for $2.5\leq D\leq 7$.}
\label{fig:estimates_allD}
}


In this section we shall collect together the information we have found about oscillons and
quasi-breathers, and show how the quasi-breather picture predicts oscillon evolution should depend on dimension.

We have already seen some specific examples of how the quasi-breather model predicts the evolution of
$\omega$ using (\ref{eq:t_estimate}), in Fig. \ref{fig:estimates_allD}
(left) we collect together
the predictions for a range of dimensions $2\half\leq D\geq 5$. As noted previously these tend
to be off by a factor of a few, overestimating the lifetime, but we can clearly see that the
oscillons in lower dimensions take much longer to reach the critical frequency. In higher dimensions
we find that the predicted lifetime is simply not long enough for the system to enter into an
oscillon phase.

We may now compare the lifetime for some specific initial condition to that of the quasi-breather oscillon model
and see how they compare in different dimensions

As was shown in \cite{Fodor:2006zs,Honda:2001xg}, in D=3 one can tune $\sigma$ to
produce oscillons with a large lifetime, but the level of fine-tuning necessary is extreme.
It is then surprising to find that in $D<3$,
oscillons living for $t\simeq 10^{6}$ are rather more generic and require little tuning
of initial conditions. The converse is true in higher dimensions, with the amount of precision
required to find a long-lived oscillon increasing with $D$. 
This is the main reason that the results
of \cite{Gleiser:2004an, Gleiser:2006te}, based on substituting an oscillon profile into the action, needed 
to be confirmed with the more detailed analysis we have presented.
Rather than trying to find the fine-tuned initial conditions which maximize the lifetime
in each dimension, we shall take the same Gaussian initial conditions ($\sigma=3.3$, $A=2$)
in the various dimensions and compare
the lifetime of the subsequent oscillon with that of the quasi-breather model. The results are plotted in Fig. 
\ref{fig:estimates_allD} (right) and show that the semi-analytic predictions are well matched by the numerical evolution.

In order to guide the eye we have overlaid a curve of the form 
\ba
\label{eq:lifeFit}
t_{\rm lifetime}=\frac{c}{(D_{\rm crit}-D)^{-a}},
\ea
with $c=1000$, $a=-1.5$ and $D_{\rm crit}=2.65$, suggesting that the
lifetime goes to infinity for some $D>0$, and possibly for 
$D>2$. While this is a possibility, the quasi-breather profile could conspire to have zero
amplitude for the radiation modes, we believe that the extrapolation of the quasi-breather results
to $D=2$ require a finite lifetime. In particular, Fig. \ref{fig:RoverE} does not suggest that the
radiation goes to zero at $D=2$. Note also that there is a kink in the curve of Fig. \ref{fig:estimates_allD} (right)
at $D\simeq 5$, after
which the lifetime suddenly drops (there are now only a handfull of
periods), where a similar fit gives $c=45$, $a=-0.7$ and $D_{\rm crit}=5.02$. 
Although the precise fitting parameters are not
essential, the change in power suggests a different regime (no oscillons) has set in.


\section{Conclusion\label{sec:conclusion}}

The interpretation of oscillons in terms of quasi-breathers gives a
simple and consistent picture of why such long-lived objects should
exist and may even be generated in a generic phase transition. 

In our setup, the quasi-breathers are the periodic solutions to the equations of
motion with minimal outgoing radiation. The presence of this radiation renders the quasi-breather unstable,
although since the radiation is minimal, presumably it is the longest
lived. Even without
excessive fine-tuning, localised lumps (oscillons) can approach and follow
the evolution of quasi-breathers, which hence must be considered
(at least) weak attractors in the space of classical solutions. 

A quasi-breather exists for each choice of frequency. Given an
initial choice of $\omega$, the configuration evolves through
a sequence of quasi-breathers, approaching a critical frequency,
corresponding to the lowest energy quasi-breather. The time is takes
to do so is well reproduced by considering the emitted radiation. As
the critical frequency is reached, the energy can no longer be
lowered, and the quasi-breather decays completely into radiation.

We performed this analysis for a range of D, the number of spatial
dimensions, in order to establish the stability/existence of oscillons
in higher dimensions. Using the quasi-breather approach we have provided
an understanding for the prescence of very long-lived solutions in
low dimensions and short-lived solutions in higher dimensions.
Although our method could not
cope with $D\leq 2$, it is tempting to conjecture from (\ref{eq:lifeFit}) that the lifetime
will diverge for some $D>0$, although the extrapolation of Fig. \ref{fig:RoverE} does not
seem to support this.
Perhaps including more radiation modes
and more hard work can settle this. While we have studied the dependence on $D$
we have not studied the dependence on $V(\phi)$. One especially interesting case would
be the sine-Gordon potential where we know there exist exactly periodic solutions
in $D=1$ \cite{Rajaraman:1982is} as well as $D=0$, which is just a particle in a well.

For $D>3$, the lifetimes become smaller and smaller, and oscillons
rush to reach the quasi-breather before decaying. For $D>5$
we were unable to find quasi-breathers with our shooting method, and it indeed seems like a
wholly new, oscillon-free, regime sets in. This is where the time a configuration takes to reach
a quasi-breather is longer than the lifetime of the quasi-breather itself, so no oscillon 
can get established.

In a 3+1-dimensional expanding Universe, assuming that the expansion
does not significantly influence the field evolution, it is
therefore natural to expect oscillons to be generated in phase
transitions or during (p)reheating after inflation, with lifetimes of
order $10^{4}$ in mass units, but probably not much longer. Scalar
fields thermalise on much longer timescales than this \cite{Berges:2002cz,Arrizabalaga:2004iw}, but
when including gauge degrees of freedom equilibraton is much
faster (see for instance \cite{Rajantie:2000nj,Skullerud:2003ki}). Oscillons may potentially delay this process if they are
produced in large numbers, and if the effects of the gauge field do not substantially change the picture.

In order to make further statements about this point, we need to look
for quasi-breathers in more realistic settings such as the Standard
Model and its extensions. In the restriction to the
$SU(2)$-Higgs model \cite{Farhi:2005rz}, an approach similar to the one
employed here should work, in spite of the complexity introduced by the
additional degrees of freedom. Work is under way to address this.


\subsection*{Acknowledgments}
P.M.S. is supported by PPARC and
A.T. is supported by PPARC Special Programme Grant {\it``Classical
Lattice Field Theory''} and we gratefully acknowledge the use of the UK
National Cosmology Supercomputer Cosmos, funded by PPARC, HEFCE and Silicon
Graphics.

\bibliographystyle{JHEP}
\bibliography{anderslit2JS}

\end{document}